\documentstyle[multicol,aps]{revtex}

\begin{document}

\title{
Unconstrained Variational Determination of the Kohn-Sham Potential}
\author{
Nikitas I. Gidopoulos
}
\address{
ISIS Facility, Rutherford Appleton Laboratory, 
Chilton, Didcot, Oxon, OX11 0QX, England, UK
}
%\date{\today}
\maketitle

\begin{abstract}
Density Functional Theory's Kohn-Sham (KS) potential emerges as the minimizing effective potential in an unconstrained variational scheme that does not involve fixing the unknown single-electron density. 
The physical content behind the virtual KS system, that of a non-interacting electronic system representing optimally the interacting one, is brought to light.  
\end{abstract}
%**********************************************************************

\begin{multicols}{2}

\section{Introduction}

Density Functional Theory \cite{hk,ks,mel,lieb,grossbook,nikreview} (DFT) has revolutionised the study of electronic structure. Its application, especially under the Kohn-Sham (KS) scheme \cite{ks,hadji}, prevails: 
in condensed matter physics DFT is, without doubt, the method of choice for ab initio calculations; in computational chemistry its popularity is broad and growing. Still, despite the success and widespread use, the physical content behind the virtual KS system is obscure and its introduction to DFT remains ad hoc. 
The KS system is a system of non-interacting electrons with the same ground state (gs) single-particle density as the system of interacting electrons of interest.
Some questions arise: 
Would one expect, {\em a priori}, this virtual system to represent the interacting system faithfully? In other words, would one expect, {\em a priori}, the KS state to be close to the interacting ground state $\Psi$? (Relative to other non-interacting states, in the same way we expect the Hartree-Fock (HF) or the Optimised Effective Potential \cite{sharp,talman,grabo} (OEP) state to be close to $\Psi$.)    
Further one may ask regarding the KS potential: Beyond the well-known theorem that the energy of the highest occupied orbital is the negative of the ionization energy 
\cite{balduz,sahni,von2}, would one expect the KS potential to have `physical reality' \cite{resta} and yield accurate single-particle excitation energies \cite{savin}, as does the OEP potential? \cite{goerling} 

These questions cause some degree of discomfort, because one is inclined to answer negatively (see however \cite{savin}), yet after several decades of successfully applying the KS method, we know that the KS scheme does have intuitional value and physical content and it cannot be a mere mathematical construct to obtain the single-particle density. It is important, and indicative of our level of understanding, that there is a way to predict it is so before doing the calculation. 

This Letter addresses this issue. It is shown that the KS potential is the optimal potential in a minimization that does not involve constraining the single-particle density. The KS scheme thus arises in a way entirely analogous to the OEP method. 

\section{Theorem}

To proceed, denote by $\Psi_j$, $E_j$ the $j$-th eigenstate, eigenvalue of the system of $N$ interacting electrons that we are interested in (with a non-degenerate gs): 
\begin{equation}
{\hat H} \, \Psi_j = E_j \, \Psi_j \, , \ j=1,2,\ldots  
\end{equation}
where the Hamiltonian which describes the system is 
\begin{equation}
{\hat H} = {\hat T} + {\hat V}^{\rm en} + {\hat V}^{\rm ee}   \, ,
\end{equation} 
${\hat T}$ is the electronic kinetic energy operator, 
${\hat V}^{\rm en}$ is the nuclear Coulomb attraction operator that binds the electrons and 
${\hat V}^{\rm ee}$ is the electron-electron Coulomb repulsion operator.

Denote also by $\Phi_{V , j}$, $E_{V , j}$, the $j$-th eigenstate and eigenvalue of a system of $N$ non-interacting electrons (${\hat V}^{\rm ee} = 0$), confined by a local potential $V({\bf r})$:  
\begin{equation}
{\hat H}_V \, \Phi_{V , j} = E_{V , j}  \, \Phi_{V , j}  \, , \ j=1,2, \ldots
\end{equation} 
where, 
\begin{equation} 
{\hat H}_V = {\hat T} + {\hat V} \, . 
\end{equation}
We shall omit the subscript $1$ from the gs, $\Psi_1 = \Psi$, $\Phi_{V , 1} = \Phi_V$, $E_1 = E$, $E_{V , 1} = E_V$. 
The subscript $V$ denotes dependence on $V({\bf r})$ which, can be almost any local potential. It is traditional in DFT to restrict the space of potentials to be $L^{3/2}+L^{\infty }$. Less restricted spaces also make sense \cite{lieb}.

Reminder of OEP \cite{sharp,talman}:
The following inequality holds:
\begin{equation} \label{oep}
\langle \Phi_V | {\hat H} | \Phi_V \rangle - E > 0 
\end{equation}
The optimized effective potential, $V^{\rm oep}$, is that potential which minimizes the lhs of (\ref{oep}). The non-interacting state $\Phi_{V^{\rm oep}}$ is expected to be close to $\Psi$ (compared with other non-interacting states), because $\Phi_{V^{\rm oep}}$ minimizes 
\begin{equation} \label{rr}
\langle \Phi | {\hat H} | \Phi \rangle - E %\ (\ge 0)
\end{equation}
in a restricted space, while $\Psi$ is the global minimizing state in the whole Hilbert space. 
Similarly, the Hartree-Fock Slater determinant, $\Phi_{\rm HF}$, which minimizes (\ref{rr}) over all Slater determinants, should be close to $\Psi$ as well.
It obviously holds that
\begin{equation}
E_{\rm OEP} \ge E_{\rm HF} > E \, ,
\end{equation}
where %$E_{\rm OEP} = \langle \Phi_{V^{\rm oep}} | {\hat H} | \Phi_{V^{\rm oep}} \rangle $ and $E_{\rm HF} = \langle \Phi_{\rm HF} | {\hat H} | \Phi_{\rm HF} \rangle $.
\begin{eqnarray}
E_{\rm OEP} & = & \langle \Phi_{V^{\rm oep}} | {\hat H} | \Phi_{V^{\rm oep}} \rangle \, ,   \\
E_{\rm HF} & = & \langle \Phi_{\rm HF} | {\hat H} | \Phi_{\rm HF} \rangle  \, .
\end{eqnarray}

Another inequality also holds, since $E_V$ is the gs energy of $H_V$:
\begin{equation} \label{m}
\langle \Psi | {\hat H}_V | \Psi \rangle - E_V > 0 \, .
\end{equation}
Ineq. \ref{m} is the dual of (\ref{oep}), as the roles of the Hamiltonians and the states over which we calculate expectation values are reversed.
%$E_V$ is a lower bound of $\langle \Psi | {\hat H}_V | \Psi \rangle$. 
We may interpret (\ref{m}), by thinking of $\Psi$ as being somewhat near the gs $\Phi_V$ of $H_V$. It makes sense to ask about the quality of this approximation and to optimize it: Keeping $\Psi$ fixed, what is the potential $V_s$ which makes the lhs of (\ref{m}) as small as possible? 

The gs, $\Phi_{V_s}$, of this optimal potential will again be close to $\Psi$ and will play a similar role to $\Phi_{V^{\rm oep}}$ or $\Phi_{\rm HF}$.
In fact $\Phi_{V_s}$, $\Phi_{V^{\rm oep}}$ are related by the duality of the variational principles (\ref{oep}, \ref{m}) they originate from. They are expected to be similarly `close' to the gs $\Psi$ and the effective non-interacting systems they describe to be equivalent. 
In a subsequent paper, we demonstrate the advantages to employ $\Phi_{V_s}$ or $\Phi_{V^{\rm oep}}$ as a starting point to approximate $\Psi$ using perturbation theory. 

It is convenient to give a name to the energy difference on the lhs of (\ref{m}): 
\begin{equation} \label{n}
T_{\Psi}[V] \doteq \langle \Psi | {\hat H}_V | \Psi \rangle - E_V 
\end{equation} 
We will assume $V_s$ is not degenerate. The case of a degenerate optimal potential $V_s$ will be discussed later on and in detail in a subsequent paper.

\vspace{0.5cm}
{\em Theorem} \\
The potential $V_s$ which minimizes $T_{\Psi}[V]$ is the Kohn-Sham (KS) potential. The ground state of $V_s$ is the KS non-interacting state having the same density as $\Psi$.

\vspace{0.5cm}
{\em Proof} \\
When we vary the potential, 
${\hat V} \rightarrow {\hat V} + \epsilon \, {\hat W}$, its gs changes,
\begin{equation}
\Phi_{ V } \rightarrow \Phi_V + \epsilon \, \Phi_V' \, ,
\end{equation}
%\begin{equation}
%\Phi_{ V + \epsilon \, W} = ( 1 + \epsilon \, {\hat G}_{V} 
%\, {\hat W} )  \Phi_V \, ,
%\end{equation}
The first-order correction $\Phi_V'$ depends on $V({\bf r})$ and $W({\bf r})$. 
For the proof, we only need that, for $\epsilon \rightarrow 0$, $\Phi_V'$ is orthogonal to $\Phi_V$: 
%\begin{equation}
%{\hat G}_{V} = - \sum_{n > 1} { | \Phi_{V , n} \rangle 
%\, \langle \Phi_{V , n} | \over E_{V , n} - E_{V } } \, .
%\end{equation} 
%Obviously, it holds
\begin{equation} \label{z}
\langle \Phi_V | \Phi_V ' \rangle = 0 %{\hat G}_{V} \, | \Phi_V \rangle = 0 \, .
\end{equation}

At the minimum, the first-order variations of $T_{\Psi}[V]$ in (\ref{n}) must vanish.
Upon varying the potential around the minimizing value $V_s$ we have: 
\begin{eqnarray}
%\lefteqn{ 
&& 
\lim_{\epsilon \rightarrow 0} \, {1 \over \epsilon} \left\{ 
\left[ 
\langle \Psi | {\hat H}_{V_s} + \epsilon \, {\hat W}| \Psi \rangle  
\right. 
\right. 
\nonumber \\
&& 
\left. 
\hspace{1.5cm} - \langle \Phi_{V_s} + \epsilon \, \Phi_{V_s} ' | {\hat H}_{V_s} 
+ \epsilon \, {\hat W}| \Phi_{V_s} + \epsilon \, \Phi_{V_s} ' \rangle 
\right] 
%} 
\nonumber \\
&& \hspace{2cm} \left. - \left[ \langle \Psi | {\hat H}_{V_s} | \Psi \rangle - E_{V_s} \right] \right\}  =  0  \, .
\end{eqnarray}
Using (\ref{z}), we obtain
%\begin{eqnarray}
%&& \ \ \langle \Psi | {\hat W}| \Psi \rangle
%- \langle \Phi_{V_s} | {\hat W}| \Phi_{V_s} \rangle \nonumber \\
%&& - E_{V_s} \, \langle \Phi_{V_s} | {\hat W} \, {\hat G}_{V_s}   
%+ {\hat G}_{V_s} \, {\hat W} | \Phi_{V_s} \rangle = 0 \, .
%\end{eqnarray}
\begin{equation}
\langle \Psi | {\hat W}| \Psi \rangle
- \langle \Phi_{V_s} | {\hat W}| \Phi_{V_s} \rangle = 0 \, .
\end{equation}
Therefore, for any $W({\bf r})$:
\begin{equation}
\int d {\bf r} \, W ({\bf r}) \, \big( \rho_{\Psi}({\bf r}) - \rho_{V_s}({\bf r}) 
\big) = 0 \, ,
\end{equation}
where $\rho_{\Psi}({\bf r})$ is the density of $\Psi$ and $\rho_{V_s}({\bf r})$ the density of $\Phi_{V_s}$.
Since the above must hold for any $W({\bf r})$, we have for all ${\bf r}$:
\begin{equation} \label{thrm}
\rho_{\Psi}({\bf r}) - \rho_{V_s}({\bf r}) = 0  \, .
\end{equation}
The lhs of the above is the functional derivative of $T_{\Psi}[V]$ at $V_s$,  
\begin{equation}
\left. { \delta T_{\Psi}[V ] \over \delta V ({\bf r}) } \right|_{V_s} = \rho_{\Psi}({\bf r}) - \rho_{V_s}({\bf r}) \, ,
\end{equation}
which must vanish.  

We have shown that $V_s$ has the same gs density as the interacting external 
potential $V^{\rm en}$. 
By definition, the KS potential is the non-interacting potential with the same gs density as the interacting one \cite{ks}. The basic theorem of DFT by Hohenberg and Kohn \cite{hk} says that we cannot have two different potentials (here both non-interacting) with the same gs density. Therefore, $V_s$ must be the KS potential and $\Phi_{V_s}$ the KS state. 

\section{Discussion}

The optimal state $\Phi_{V_s}$ need not be a single Slater determinant. We may require that it satisfies symmetries of ${\hat H}$, for example we may choose it an eigenstate of $S^2$ and $S_z$.

The value of $T_{\Psi}[V]$ at the minimum is
\begin{equation}
T_{\Psi}[V_s] = \langle \Psi | {\hat T} | \Psi \rangle - 
\langle \Phi_{V_s} | {\hat T} | \Phi_{V_s} \rangle > 0 \, .
\end{equation}
One may recognise $T_{\Psi}[V_s]$ as the kinetic part, $T_{\rm c}[\rho_\Psi]$, of the correlation energy functional $E_{\rm c}[\rho_\Psi]$ 
$
= \langle \Psi | {\hat T} + {\hat V}^{\rm ee} | \Psi \rangle - 
\langle \Phi_{V_s} | {\hat T} + {\hat V}^{\rm ee} | \Phi_{V_s} \rangle
$.

The optimal potential $V_s$ is determined with a freedom of a constant.
Separately, the energies $\langle \Psi | {\hat H}_{V_s} | \Psi \rangle$ 
and $E_{V_s}$ appearing in 
$T_{\Psi}[V_s]$ do not give any estimate of the total energy, $E$, since they are arbitrary within a constant shift. 

The expectation value $\langle \Phi_{V_s} | {\hat H} | \Phi_{V_s} \rangle$ gives an upper bound to $E_{\rm OEP}$,  
\begin{equation}
\langle \Phi_{V_s} | {\hat H} | \Phi_{V_s} \rangle 
\ge 
E_{\rm OEP}
\end{equation}
because $V^{\rm oep}$ minimizes (\ref{oep}).

In the derivation of (\ref{thrm}) we never used that $\Psi$ is the gs of $\hat H$. 
$\Psi$ could be any state and then the minimization of $T_{\Psi}[V]$ would 
yield a non-interacting potential $V_s$ whose gs $\Phi_{V_s}$ would have the same single-particle density as $\Psi$. 
The question of non-interacting $v$-representability arises here non-trivially: 
%Rigorously, we do not know whether the minimum of $T_\Psi[V]$ exists for all $\Psi$.
%Only the infimum can be strictly defined. Even if the minimum of $T_\Psi[V]$ did 
%exist, we would not be sure that the derivative at the minimum is zero. 
%In either case (\ref{thrm}) might not hold. 
does the optimal $V_s$ exist in the space of potentials considered? 
Going back to the proof of the theorem, rigorously, we do not know whether the minimum of $T_\Psi[V]$ exists for all $\Psi$. 
Only the infimum can always be defined and then (\ref{thrm}) might not hold exactly. 
However, having an {\it unconstrained variational principle} (\ref{m}) to determine the optimal potential, guarantees that 
%by constructing a steepest-descent algorithm, 
we may approach numerically the infimum of $T_\Psi[V]$ as close as we wish. 

An example will illustrate this: $\Psi$ can be chosen an excited state of ${\hat H}$. Then, the non-interacting potential $V_s$ should develop divergences to simulate in its gs density the nodes present in the excited state density $\rho_{\Psi}$.
We may construct a steepest-descent algorithm to minimize $T_{\Psi}[V]$.  
This algorithm would generate a series of non-interacting potentials with spikes of increasing height at the nodes of $\rho_{\Psi}$, so that the gs densities of these potentials would approach $\rho_{\Psi}$ successively closer.

A similar variational principle to (\ref{m}) can be used to derive the optimal effective spin-up/down potential when the interacting system lies in a weak uni-axial magnetic field defining the $z$ direction. Here, we still denote the gs of the interacting system by $\Psi$. 
Even in the absence of an external magnetic field, allowing for different effective spin-up/down potentials $V \rightarrow (V^\uparrow , V^\downarrow)$ in the energy difference $T_{\Psi}[V]$ in (\ref{n}), in general, will lead after optimization to a lower minimum, thus implying that the non-interacting state $\Phi_{(V^\uparrow_s , V^\downarrow_s)}$ offers a better representation of $\Psi$ than the original 
state $\Phi_{V_s}$. Physically, we expect this to become relevant when the electrons in the system do not form closed shells.  
The variational principle that after minimization will determine the spin-potential is: \begin{equation} \label{nup}
%T_{\Psi}[V^\uparrow , V^\downarrow ] \doteq  
\langle \Psi | {\hat H}_{( V^{\uparrow }, V^{\downarrow } )} | \Psi \rangle - E_{(V^{\uparrow }, V^{\downarrow })} > 0 \, ,
\end{equation} 
where $E_{(V^{\uparrow }, V^{\downarrow })}$ is the gs of the non-interacting spin-diagonal Hamiltonian 
\begin{equation} \label{spinh}
{\hat H}_{(V^{\uparrow }, V^{\downarrow })} = {\hat T} + {\hat V}^{\uparrow } + {\hat V}^{\downarrow } \, .
\end{equation}    
Optimizing separately the lhs of (\ref{nup}) for the spin-up and down potentials, one finds that the optimal spin-potential $(V_s^{\uparrow }, V_s^{\downarrow })$ is the KS spin-potential with the correct spin-density (unrestricted KS scheme, or KS scheme in Spin-DFT \cite{von,raj}): 
\begin{equation}
( \rho_{V^{\uparrow }}^{\uparrow } ({\bf r}) , \rho_{V^{\downarrow }}^{\downarrow }  ({\bf r}) ) =
( \rho_{\Psi}^{\uparrow }  ({\bf r}) , \rho_{\Psi}^{\downarrow }  ({\bf r}) ) \, .
\end{equation}
In the absence of an external magnetic field, since the minimum of $T_{\Psi}[V^\uparrow ,V^\downarrow ]$ (lhs of (\ref{nup})) is in general lower than the minimum of $T_{\Psi}[V]$, the following inequality holds rigorously for the kinetic part of the correlation energy functional:
\begin{equation}
T_{\rm c}[\rho_\Psi^\uparrow , \rho_\Psi^\downarrow] \le T_{\rm c}[\rho_\Psi^\uparrow +\rho_\Psi^\downarrow] \, .
\end{equation}
A similar inequality should hold, at least approximately, for the magnitude of the correlation energy functionals.

We note that a generalised HK theorem holds \cite{eschrig,nik} for the spin-diagonal Hamiltonian (\ref{spinh}), as long as spin-up and down potentials are defined with a freedom of a spin-constant and the system described by 
${\hat H}_{(V^{\uparrow }, V^{\downarrow })}$ is not perfectly spin-polarized.

Finally, let us consider the degenerate case. When the optimal potential $V_s$ is degenerate, the gs $\Phi_{V_s}$ is not  unique. Then, a non-interacting state in the space of degeneracy will have single-particle density equal to $\rho_\Psi$.
I do not find this statement very helpful though. Often degeneracy is due to open shells and unpaired electrons. Then, allowing for different spin-up/down effective potentials may lift the degeneracy. If the degeneracy persists even in the presence of an effective magnetic field (when $V_s^{\uparrow } \ne V_s^{\downarrow }$), then use of subspace or ensemble KS scheme for degenerate or low excited states \cite{theo,theo2,gok1,gok2,ghost} is appropriate. We shall devote another paper to the quasi-degenerate case and the development of an ab-initio, multi-configurational KS scheme.

In conclusion, we have shown that the KS system, until now an ad hoc construction with rather esoteric and not a priori transparent physical content, can be seen as an optimal non-interacting system to represent the gs of the interacting electronic system, completely analogous or dual to the OEP system.
This provides a bridge between wave-function methods traditionally employed in 
quantum chemistry and DFT. For example, using accurate gs wave-functions ${\tilde \Psi}$ calculated by other methods and minimizing the energy difference functional $T_{\tilde \Psi}[V]$, we can obtain accurate KS potentials for the studied systems and assess the accuracy of existing approximate  exchange-correlation potentials \cite{robert,umrigar,morrison,wu}.
More importantly, the prospect to explore and devise new expressions for the correlation energy and potential based on the variational principle (\ref{m}) appears promising.

K. Burke, R. van Leeuwen, M.W. Long and C.A. Ullrich communicated helpful comments on an early version of this manuscript.

\end{multicols}

\end{document}